\documentclass[]{aastex631}

\usepackage{amsmath}
\usepackage{float}
\usepackage{graphicx}
\usepackage{soul}
\usepackage{ulem}

\begin{document}

\title{GRB 080503: A very early blue kilonova and an adjacent non-thermal radiation component}

\author{Hao Zhou}
\affiliation{Key Laboratory of Dark Matter and Space Astronomy, Purple Mountain Observatory, \\
Chinese Academy of Sciences, Nanjing 210023, China}
\affiliation{School of Astronomy and Space Science, University of Science and Technology of China, \\
Hefei 230026, China}

\author{Zhi-Ping Jin}
\affiliation{Key Laboratory of Dark Matter and Space Astronomy, Purple Mountain Observatory, \\
Chinese Academy of Sciences, Nanjing 210023, China}
\affiliation{School of Astronomy and Space Science, University of Science and Technology of China, \\
Hefei 230026, China}

\author{Stefano Covino}
\affiliation{INAF/Brera Astronomical Observatory, via Bianchi 46, I-23807 Merate (LC), Italy}

\author{Lei Lei}
\affiliation{Key Laboratory of Dark Matter and Space Astronomy, Purple Mountain Observatory, \\
Chinese Academy of Sciences, Nanjing 210023, China}
\affiliation{School of Astronomy and Space Science, University of Science and Technology of China, \\
Hefei 230026, China}

\author{Yu An}
\affiliation{Key Laboratory of Dark Matter and Space Astronomy, Purple Mountain Observatory, \\
Chinese Academy of Sciences, Nanjing 210023, China}
\affiliation{School of Astronomy and Space Science, University of Science and Technology of China, \\
Hefei 230026, China}

\author{Hong-Yu Gong}
\affiliation{Key Laboratory of Dark Matter and Space Astronomy, Purple Mountain Observatory, \\
Chinese Academy of Sciences, Nanjing 210023, China}
\affiliation{School of Astronomy and Space Science, University of Science and Technology of China, \\
Hefei 230026, China}

\author{Yi-Zhong Fan}
\affiliation{Key Laboratory of Dark Matter and Space Astronomy, Purple Mountain Observatory, \\
Chinese Academy of Sciences, Nanjing 210023, China}
\affiliation{School of Astronomy and Space Science, University of Science and Technology of China, \\
Hefei 230026, China}

\author{Da-Ming Wei}
\affiliation{Key Laboratory of Dark Matter and Space Astronomy, Purple Mountain Observatory, \\
Chinese Academy of Sciences, Nanjing 210023, China}
\affiliation{School of Astronomy and Space Science, University of Science and Technology of China, \\
Hefei 230026, China}

\correspondingauthor{Zhi-Ping Jin, Yi-Zhong Fan}
\email{jin@pmo.ac.cn, yzfan@pmo.ac.cn}

\begin{abstract}
The temporal behavior of the very dim optical afterglow of GRB 080503 is at odds with the regular forward shock afterglow model and a sole kilonova component responsible for optical emission has been speculated in some literature. Here we analyze the optical afterglow data available in archive and construct time-resolved spectra. The significant detection by Keck-I in {\it G/R} bands at $t\sim 3$ day, which has not been reported before, as well as the simultaneous Gemini-North {\it r} band measurement, are in favor of a power-law spectrum that is well consistent with the optical to X-ray spectrum measured at $t\sim 4.5$ day. However, for $t\leq 2$ day, the spectra  are thermal-like and a straightforward interpretation is a kilonova emission from a neutron star merger, making it, possibly, the first detection of a very early kilonova signal at $t\sim 0.05$ day. A non-thermal nature of optical emission at late times ($t\sim 2$ day), anyhow, can not be ruled out because of the large uncertainty of the {\it g}-band data. We also propose to classify the neutron star merger induced optical transients, according to the temporal behaviors of the kilonova and the non-thermal afterglow emission, into four types. GRB 080503 would then represent the first observation of a sub-group of neutron star merger driven optical transients (i.e., Type IV) consisting of an early blue kilonova and an adjacent non-thermal afterglow radiation.
\end{abstract}


\section{Introduction}\label{Sec:Introduction}
In addition to the strong gravitational wave radiation, binary neutron star (BNS) mergers can lead to plentiful electromagnetic phenomena, such as the short/hybrid Gamma-Ray Bursts (GRBs) generated by the internal energy dissipation of the narrowly-collimated relativistic outflow, the subsequent forward shock afterglow driven by the interaction between the GRB ejecta and the interstellar medium, and the kilonova/macronova radiation from the radioactive decay of heavy material synthesized in the sub-relativistic  outflow launched by the merger \citep{1989Natur.340..126E,1998ApJ...507L..59L,2019LRR....23....1M}, as convincingly observed in GW170817/GRB 170817A/AT2017gfo \citep{2017PhRvL.119p1101A,2017ApJ...848L..14G,2017Natur.551...64A,2017Natur.551...67P}. Before the launch of the gravitational wave astronomy era, kilonova candidates (i.e., the thermal like emission in near-infrared/optical/ultraviolet bands, depending on the lanthanide composition of the emitting region) have been identified in GRB 130603B \citep{2013Natur.500..547T,2013ApJ...774L..23B}, GRB 060614 \citep{2015NatCo...6.7323Y,2015ApJ...811L..22J} and GRB 050709 \citep{2016NatCo...712898J}. Later on, kilonova candidates have also been reported in GRB 160821B \citep{2018ApJ...857..128J,2019MNRAS.489.2104T,2019ApJ...883...48L}, GRB 150101B \citep{2018NatCo...9.4089T}, GRB 070809 \citep{2020NatAs...4...77J} and GRB 060505 \citep{2021arXiv210907694J}.  

\begin{figure}[!h]
\centering
\includegraphics[width=0.95\columnwidth]{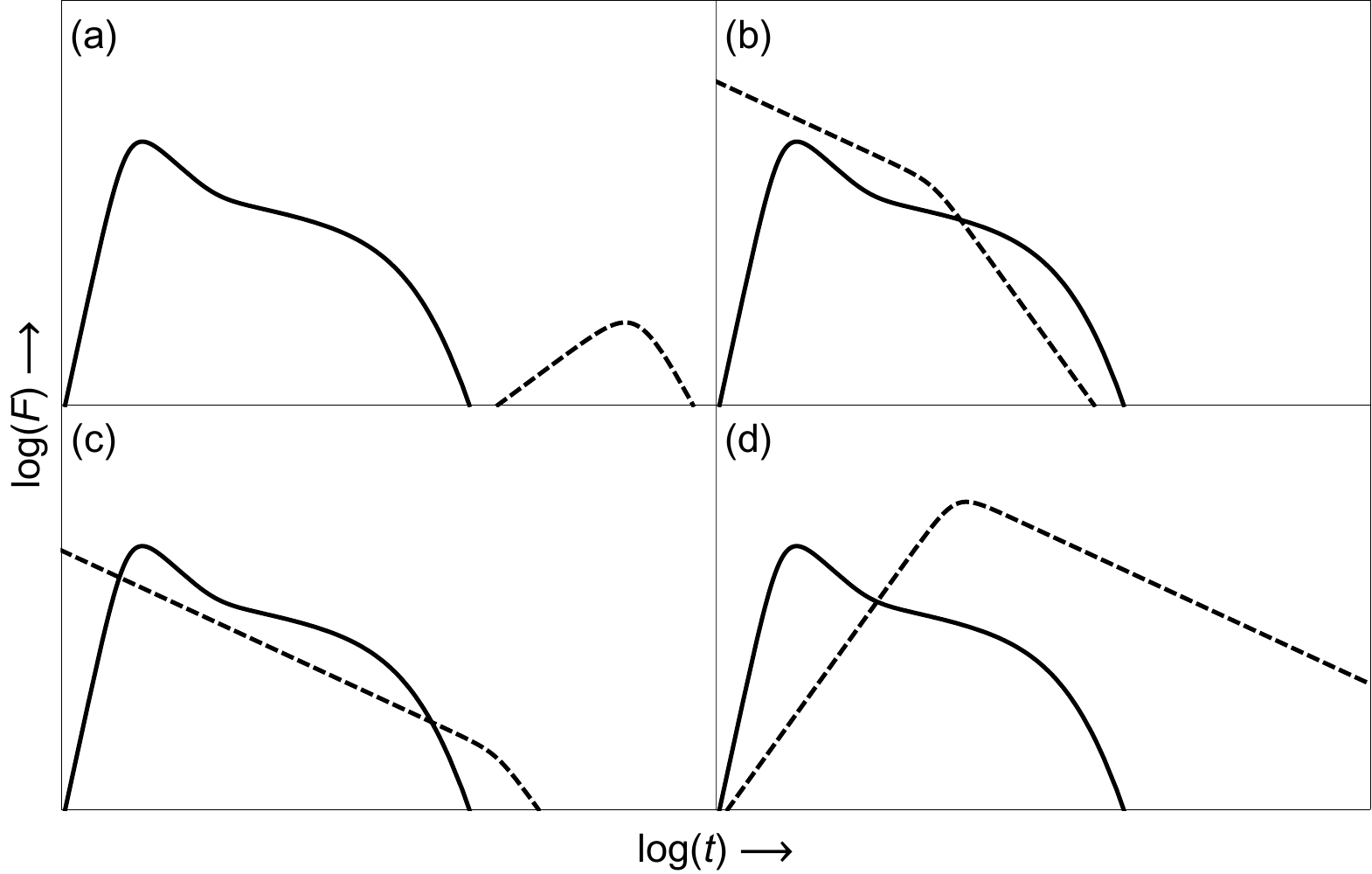}
\caption{{\bf Synthetic cartoon ultraviolet/optical/near-infrared emission light curve of neutron star merger events based on the current observational data of GRBs.} The solid line and the dashed line represent the kilonova component and the forward shock afterglow radiation, respectively. {Here, for simplicity we do not include the possible reverse shock emission as well as the radiation from the prolonged activity of the central engine.} Note that  late time kilonova emission (i.e., the ``bump"-like component) is mainly in infrared band, which is usually challenging to observe for ground-based telescopes unless the source is nearby. 
GRB170817A/AT2017gfo, GRB 130603B, GRB 060505 are representatives of Type I (i.e. (a)), Type II (i.e. (b)) and Type III (i.e. (c)) classes, respectively. Type IV (i.e. (d)) is a new class observed likely for the first time in GRB 080503.
}\label{fig:Cartoon}
\end{figure}

Based on the temporal behaviors of optical/infrared afterglows and the kilonova (including also the candidates) radiation, the current sample can be divided into three groups (see the schematic plots in Fig.\ref{fig:Cartoon} for a summary).
The first group (hereafter Type I) is represented by GRB 170817A/AT2017gfo and possibly also GRB 150101B that are characterized by the ``early" kilonova emission followed by a well-separated long-lasting afterglow at late times (i.e., the peaks of these two components are well separated). 
In the future, Type I is likely to be common because the energetic core of the GRB ejecta are typically characterized by a half-opening angle of $\sim 0.1$ rad and therefore compact object mergers identified by their GW emission will be mainly viewed highly off-axis. Their afterglow emission can only be detected at very late time when the bulk Lorentz factor of the outflow has dropped significantly. The second group (hereafter Type II) consists of GRB 130603B, GRB 060614 and GRB 050709, which are characterized by the emergent kilonova radiation in the quick decline phase (i.e., due to the jet effect) of the forward shock emission. Type II dominates the current sample because most of the current short/hybrid GRBs were detected at a redshift of $\geq 0.1$, for which the GRBs were viewed on-axis otherwise too dim to be reliably detected. The third group (i.e., Type III) is featured by a luminous blue kilonova superposed on the forward shock emission which is still in the normal decline phase (i.e., before the so-called jet break). GRB 060505 and GRB 070809 likely belong to this group (though for GRB 070809, the forward shock optical emission was just indirectly inferred because of the scarcity of the data). GRB 160821B belongs to either Type II \citep{2019ApJ...883...48L} or Type III \citep{2019MNRAS.489.2104T}, depending on the ``modeling” of the forward shock radiation. There could be the fourth group (hereafter Type IV) that is dominated by the kilonova emission at early time, but then a non-thermal optical emission, e.g. forward shock, becomes brighter than the kilonova in very short duration and dominates the temporal behavior at late time.

In this work we report a careful re-analysis of the GRB 080503 and show that it likely represents the first observation of a Type IV class event.

GRB 080503 was discovered by the {\it Swift} satellite and its X-ray emission was detected at $t\leq 10^{3}$ sec \citep{2008GCN..7665....1M,2008GCN..7669....1G}. The prompt emission light curve of GRB 080503 might consist of two parts, including an initial spike with 15-150 keV $T_{90}$ duration of 0.32$\pm$0.07 s and the much longer extended emission. The ratio of the fluence of extended emission to that of the initial spike is $\sim$ 32, which makes GRB 080503 an outlier of short GRBs with extended emission (note that even for GRB 060614, the outstanding/famous hybrid burst, this ratio is just $\sim 6$ \citep{2009ApJ...696..971X}). The lack of host galaxy down to the limit of 28.5th magnitude by Hubble Space Telescope (HST) favors a merger origin (i.e., it is a short burst \citep{2009ApJ...696.1871P}). Anyhow, the lack of a reliable measurement of the redshift of this event as well as the long temporal lag of the extended emission render the situation less clear \citep{2009ApJ...696..971X}.  
Dedicated follow-up optical observations were carried out by Gemini-N, Keck-I and Hubble Space Telescope (HST). Surprisingly, at $t\sim 0.05$ day after the burst the optical emission was down to $\sim 26$th magnitude (in this work all of the magnitudes are measured in the AB system) and got brightened to $\sim 25$th magnitude at $t\sim 1$ day \citep{2009ApJ...696.1871P}. Such behaviors are quite different from the regular forward shock afterglow model prediction, in which the optical emission usually peak at a time $\leq 10^{3}$ s unless the bulk Lorentz factor of GRB ejecta is much lower than the typical value of $\sim 10^{2}-10^{3}$ \citep{2004RvMP...76.1143P}. One possibility is that the rebrightening is the emergence of a Li-Paczynski kilonova/macronova \citep{1998ApJ...507L..59L,2005astro.ph.10256K}, as hypothesized in Perley et al. \citep{2009ApJ...696.1871P}. However, at such an early time, the hypothesis was motivated by the unusual temporal behavior of the optical emission and the calculation of the macronova emission was simply attributed to the Nickel decay. Further examinations on the origin of the optical emission were carried out in the literature. Some colleagues argued that the panchromatic rebrightening was caused by a refreshed shock \citep{2012A&A...541A..88H}. A kilonova/macronova with a relatively ``long-lived" ($\sim 100$ ms) hypermassive neutron star as its central engine was also studied and the $r$-band data could be reasonably fitted according to this scenario \citep{2015MNRAS.450.1777K}. Some colleagues argued that the $r-$band data were powered by a long-living magnetar \citep{2015ApJ...807..163G,2017MNRAS.470.4925G}. Clearly, all these late investigations simply adopt the data reported initially in \cite{2009ApJ...696.1871P} and all works \citep[including][]{2009ApJ...696.1871P} did not pay much attention to the spectral properties, due to the sparsity of the data. However, the detailed spectral analysis is found to be essential in identifying the kilonova signal, as demonstrated for instance in GRB 060505 \citep{2021arXiv210907694J}. Therefore, in this work we re-analyze all the afterglow data of GRB 080503 to search for the spectral evidence of the presence of a kilonova and then examine whether there is an unambiguous signal for a non-thermal optical component. Note that throughout this work we adopted cosmological parameters from Planck Collaboration \citep{2020A&A...641A...6P}, $\Omega_{\rm m}=0.315\pm0.007$ and $H_{\rm 0}=(67.4\pm0.5)$ km s$^{-1}$ Mpc$^{-1}$.

\begin{figure}[!h]
\centering
\includegraphics[width=0.495\columnwidth]{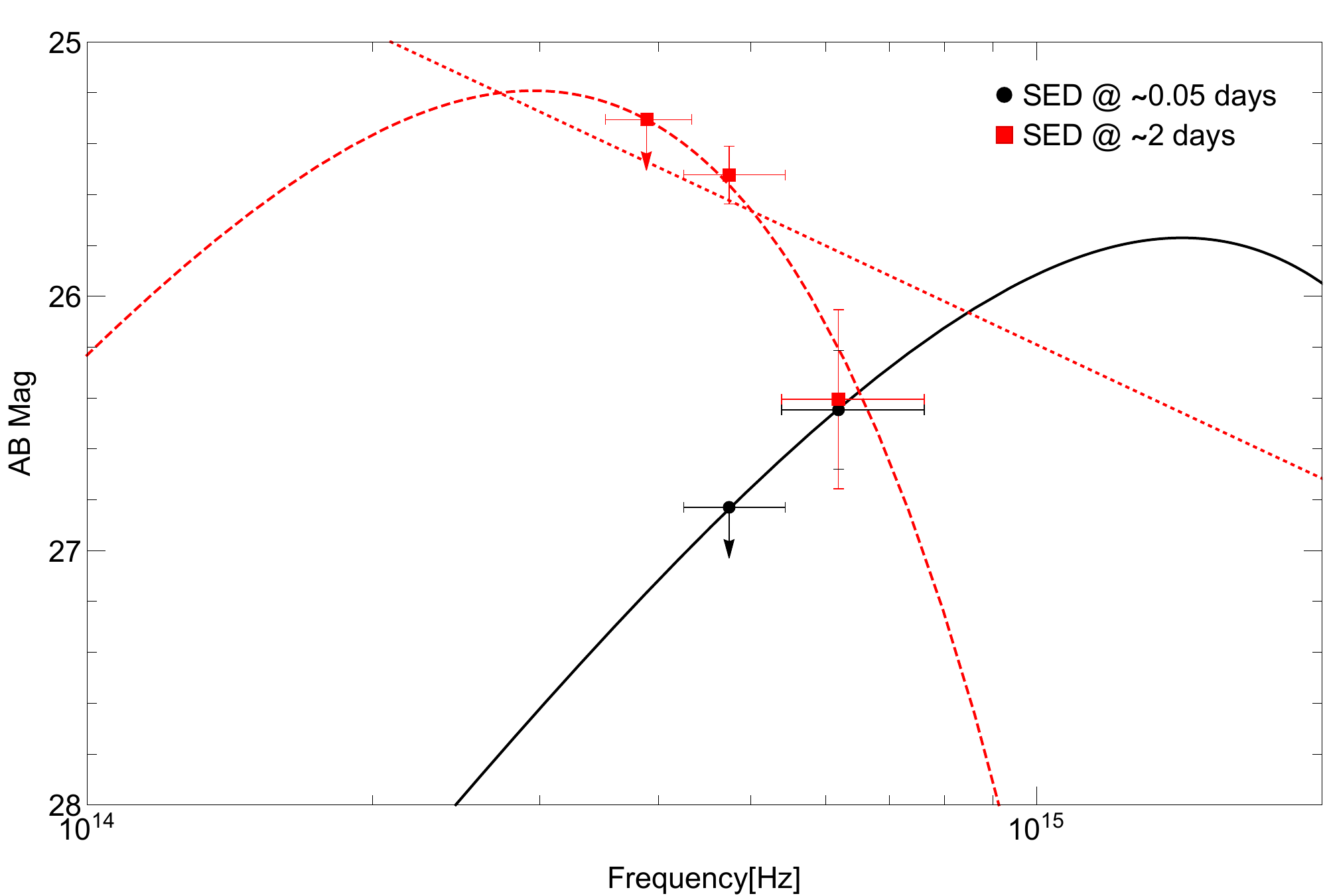}
\includegraphics[width=0.495\columnwidth]{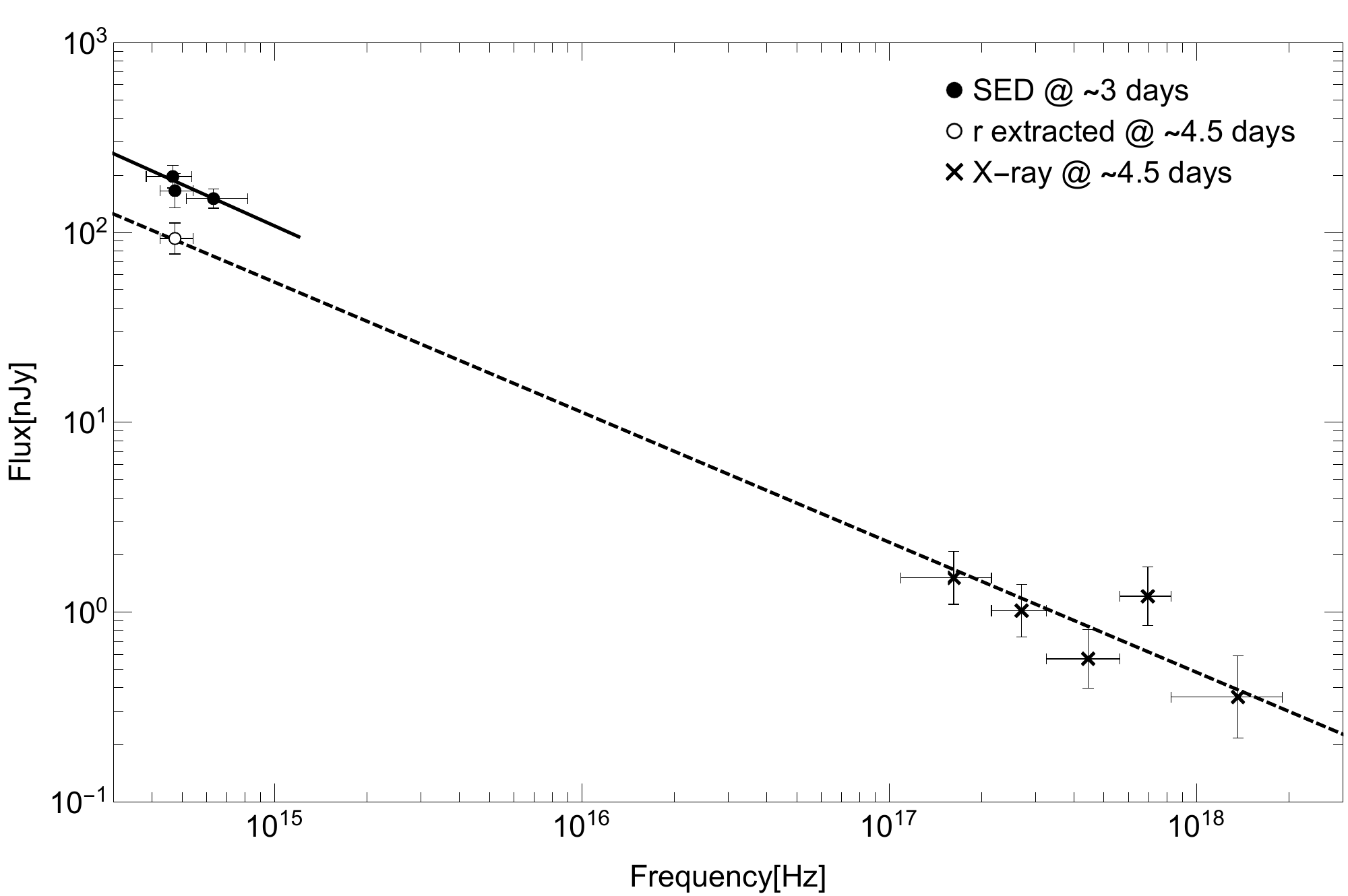}
\caption{{\bf Time-resolved spectral energy densities (SEDs) of the optical (and X-ray) emission of  GRB 080503.} The left panel shows the SEDs of the optical emission at $t=(0.05,~2)$ day. At $t=0.05$ day, the spectrum is very hard and indicates a thermal radiation at the high temperature of $\sim 2\times 10^{4}$ K. The  radiation at $t=2$ day can be fitted by a thermal spectrum, too. However, the uncertainties of these data points are large and a power-law spectrum of $f_\nu \propto \nu^{-0.7}$ (i.e., the red dotted line, as motivated by the late time spectra presented in the right panel) can not be convincingly ruled out. The right panel shows the SEDs constructed at $t=(3,~4.5)$ day. For clarification, some other much-less stringent upper limits are not shown. Slopes of solid line and dashed line are $-0.73\pm0.52$ and $-0.69\pm0.04$, respectively. These two spectral slopes are consistent with each other, revealing the non-thermal nature of the radiation. The X-ray data points have been corrected for the absorption of Milky Way as well as the ``host" (the latter is found to be negligible, see Table \ref{tab:X-ray_Fitting}). The optical data is only corrected for the extinction of Milky Way, since there is no host galaxy identified at the site of GRB 080503.}
\label{fig:SED005}
\end{figure}

\section{Observations}
\subsection{General Optical Data Analysis}
Here we report procedures applied for optical data analysis and focus on the details of some new data or the difference in comparison to that found in \cite{2009ApJ...696.1871P}. Our results are summarized in Table \ref{tab:080503Phot}.
Full Width at Half Maximums(FWHMs) of Gemini-North and Keck-I images are similar $\sim0.7$ arcsec, which corresponds to $\sim$ 5 pixels for both telescopes, because of the similarity of their pixel scales (i.e., $\sim$ 0.14 arcsec/pixel). Typically, we use circles with radii ranging from 2 to 18 pixels as source regions and an annulus with an inner radius of 20 pixels and an outer radius of 30 pixels as background region, and get a curve of magnitude versus aperture radius (M-R curve in short). Unsaturated stars near the source region are selected carefully to measure the growth curve for each image, i.e. the stars should be bright enough and isolated. Data points from the M-R curve ranging from 2 to 5 pixels are fitted by the growth curve to make aperture correction. We made photometry with the \textit{Photutils} \citep{larry_bradley_2022_6825092} package, which is not only able to build a 2D sky background but also a 2D uncertainty map for sky background. Uncertainties of measurements were estimated with it as well.

For images those do not show significant signals (i.e., ${\rm SNR}<$3) at the position of transient, we estimate the fluctuation of photon counts in background region (the annulus mentioned above) and calculate how much should the flux of transient be to produce a signal with ${\rm SNR}\sim 3$. The SNR is defined as ${\rm SNR} = {S}/{\sqrt{S+A_{S}\sigma _{b}^{2}(1+A_{S}/A_{B})}}$, where $A_{S}$ and $A_{B}$ represent areas of source and background regions, respectively. Photoelectron counts from source (subtract background value from the total counts in source region) $S$ is treated as obeying Poisson distribution, hence the variance of $S$ is itself, and $\sigma_{b}$ is the fluctuation of background. The factor $1+A_{S}/A_{B}$ has practical meanings: If $A_{S}=A_{B}$, the factor of $A_{S}\sigma_{b}^{2}$ is 2, which means the background is calculated twice in this measuring method. For a infinite background region, the effects of background in source region is negligible, while under the case of $A_{S}\gtrsim A_{B}$, one should be cautious about whether the mean background value and fluctuation derived from such small region are properly estimated.

\subsection{Gemini-North Observations}
About 1 hour after the burst, Gemini-N observed GRB 080503 in order of $r$, $g$, $r$, $i$, $z$ and $g$ bands, and 5$\times$180 s exposure time for all exposures except the first one, for which is 180 s \footnote{PI: Joshua Bloom}. Since the sky was brightening rapidly, only the first $g$-band image revealed a new source with a signal to noise ratio (SNR) $>3$. In Table \ref{tab:080503Phot}, the two early $r$-band measurements have been combined to set a stringent upper bound. On May 5, 2008, Gemini-N took a 9$\times$180s exposure in $r$ band and a 4$\times$180 s exposure in both $g$ and $i$ bands. On May 4, 6 and 7, 2008, Gemini-N took 10$\times$180 s, 15$\times$180 s and 16$\times$180 s exposures in $r$ band, respectively.

Landolt $UBVR_{\rm C}I_{\rm C}$ magnitudes of standard field SA110-361 \citep{1992AJ....104..340L} were transformed to $ugriz$ with the equations of \cite{2002AJ....123.2121S}, similar to the calibration document of GMOS/Gemini-N \citep{2009PASA...26...17J}. We found that the $r$-band image quality of SA110-361 field taken on May 5 is poor, hence the zeropoint is not reliable as those derived from other days. As a result, the zeropoint derived from image taken on May 4 was used to calibrate $r$-band images taken on May 5, 6 and 7. In addition, we selected 12 galaxies in the field of Gemini-N to check zeropoint drifts, and photometries of 12 galaxies indicate that $i$-band photometry of May 5 and $r$-band photometry of May 6 should be brightened by $0.036\pm0.0285$ and $0.195\pm0.0200$ mag, respectively.

In general, our analysis results are well consistent with that reported in \cite{2009ApJ...696.1871P} except that for the $g$-band measurement on 5 May since our error bar is about 1.35 times of the value reported before. We have checked uncertainties caused by each step, which are $\sim 0.30$ mag, $\sim 0.02$ mag, $\sim 0.15$ mag and $\sim 0.12$ mag for source measurement, growth curve measurement, aperture correction and zeropoint. All of these terms yield a total uncertainty of $\sim 0.35$ mag. We have also analyzed the almost simultaneous $i-$band observation and derived an upper limit $>$ 25.3th mag to bound the spectral shape.

\subsection{Keck-I Observations}
Keck-I telescope observed the GRB 080503 field at $\sim$ 0.05 day after BAT's trigger with LRIS in $B$ and $R$ bands simultaneously \footnote{PI: Mike Bolte}. The total exposure time in $B$ band is 690 s (1$\times$30 s, 1$\times$60s and 2$\times$300s). However, after a careful examination, we found that the transient is outside the field of view for the 2$\times$300 s exposures, hence the effective exposure time for $B$ band is only 90s, which is too short to derive a stringent constraint. That is why our $B$-band upper limit is much looser than that presented in \cite{2009ApJ...696.1871P}. In $R$ band, 1$\times$30 s and 2$\times$300 s 
exposures were taken. 
Only the 2$\times$300 s exposures were stacked for our scientific measurement. This is due to the background fluctuation of the 1$\times$30 s exposure that is much higher than that of the 2$\times$300 s exposures, and a poorer constraint would be set if we stack all 3 exposures.

On May 6, 2008, Keck-I observed GRB 080503 field again with LRIS in $G$ and $R$ bands\footnote{PI: Wallace Sargent}. Each band took 6$\times$1200s exposures and the optical transient was successfully detected in both bands. Since the gain(i.e., $e^-$/ADU) of the first exposure in $G$ band is much larger than that of other five exposures, the last five exposures are combined. It is the same for the $R$-band images. Both of $G$ and $R$ detection on May 6 are reported for the first time, and then provide crucial information on the optical afterglow spectrum of GRB 080503. 

The photometry method for Keck images is almost same as that for Gemini images, but background thresholds is slightly different.
Only signals with S/N ratio larger than 5 are treated as true signals, because image quality of LRIS is not as good as GMOS.

Since there is no standard star field taken by Keck-I, we carefully selected some secondary standard stars (unsaturated and isolated but not too faint) for each image taken by Keck to get as good as possible zeropoint calibration. These stars are listed in Table \ref{tab:GandKMay3} and \ref{tab:GandKMay6}. Transformation equations in \cite{2002AJ....123.2121S} give Vega $B$ and $R$ magnitudes, and to get AB magnitudes, they should be added by -0.128 and 0.178, respectively.

\subsection{HST Observations}
Hubble Space Telescope(HST) observed GRB 080503 field with Wide Field Planetary Camera 2(WFPC2) on May 8, May 12 and July 29, 2008\footnote{PI: Joshua Bloom}. All the HST data used in this paper can be found in MAST: \dataset[https://doi.org/10.17909/gepa-1j39]{https://doi.org/10.17909/gepa-1j39}. The first observation took a 2100s exposure in each of F450W and F814W band, and a 4600s exposure in F606W band. The second observation took a 4000s exposure in each of F606W and F814W band. The last observation took a 9200s exposure in F606W band.

The optical transient is only successfully detected in F606W-image taken on May 8($\sim27.0$ mag). The F606W-image taken on July 29 does not show any galaxy at position of the transient, but there is a very faint galaxy $\sim0.8''$ from the transient's position with F606W magnitude $\sim27.3$ mag \citep{2009ApJ...696.1871P}. The galaxy is fainter than the detection limit of Gemini and Keck, so it has no influence on photometry results derived from Gemini and Keck.

\subsection{Other Optical Observations}
Just $\sim$ 2 minutes after the BAT trigger, the Ultraviolet Optical Telescope(UVOT) on \textit{Swift} observed the GRB 080503 field in White band. On May 8, NIRI at Gemini-North also observed the transient in $K_{s}$ band\footnote{PI: Joshua Bloom}.
However, comparing with the photometry reported in Table \ref{tab:080503Phot}, both observations just give very poor constraints for transient behavior\citep{2009ApJ...696.1871P}, hence we do not list their results here.

\subsection{X-ray Data Analysis}
The GRB 080503 was observed several times by the Neil Gehrels $Swift$ Observatory X-Ray Telescope (XRT). However, significant detection was only achieved in the early $\sim 2000$ s \citep{2009ApJ...696.1871P}. Such early time X-ray data is not helpful in constructing the late time wide band afterglow SED. Below we focus on the observations of $Chandra$ X-Ray Observatory. 

GRB 080503 was observed twice by the $Chandra$ X-Ray Observatory. We used the first $Chandra$ observation (i.e., observation identifier 9853) to extract the spectrum and flux. After reducing the data with $CIAO$ pipeline, we detected the GRB 080503 source position at $RA$=19$^{\rm h}$06$^{\rm m}$28.75$^{\rm s}$, $DEC$=+68$^{\circ}$47'35.39" above 5$\sigma$ with the $wavdetect$ command of $CIAO$ software in version 4.13. We extract spectrum in two circle regions with radius of 0.3" (3.6 pixels) and 4" (44 pixels) for the source and background respectively.

The  first $Chandra$ observation of GRB 080503 is during 4.29 $\sim$ 4.66 days after $Swift$ trigger time. As usual, the intrinsic spectrum is assumed to be a single power-law and the observed emission suffered from the absorption of the host galaxy as well as Milky Way. The equivalent Hydrogen column density of Milky Way is taken as $N_{\rm H}=6.99\times10^{20}$ cm$^{-2}$ (see https://www.swift.ac.uk/xrt\_spectra/00310785/) and the one in the host galaxy is taken as free. The best-fitting spectral index of $Chandra$ X-Ray spectrum is $\beta=-0.73\pm 0.27$ (68\% confidence level) under the $Cash$ statistics and the un-absorbed $0.3-10$ keV flux is $1.22\times10^{-14} {\rm erg~cm^{-2}~s^{-1}}$, as listed in Table \ref{tab:X-ray_Fitting}. The intrinsic(un-absorbed) spectrum of GRB 080503 is listed in Table \ref{tab:X-ray_Spec}.

\section{Data Analysis}\label{Sec:RESULTS}
Our re-analyzed optical observation results of GRB 080503 are summarized in Table \ref{tab:080503Phot}. Note that in \cite{2009ApJ...696.1871P}, except at $t\approx 2$ day after the burst, all the optical emission were just measured in a single band, which hampers a reliable examination on the spectral evolution. Fortunately, in the archive we noticed simultaneous observation of this source by Keck-I in $G$ and $R$ bands. The $\sim 6000$ s exposure is long enough to yield the accurate measurements. The Keck-I $R$ band flux is well consistent with the almost simultaneous measurement by Gemini-N in $r$ band. With the Keck-I and Gemini-N measurements at $t\approx 3$ day, we obtained a power-law spectrum of $f_\nu \propto \nu^{-0.73\pm 0.52}$ (see the right panel of Fig.\ref{fig:SED005}), note that the relatively large error is introduced by the short range of the frequencies. This value is nicely in agreement with the spectrum of $f_\nu \propto \nu^{-0.69\pm0.04}$ constructed with Chandra X-ray measurement at $t\approx 4.5$ d after the GRB trigger and the extrapolated $r$-band emission (see the right panel of Fig.\ref{fig:SED005}). Our current finding based on the Keck-I/Gemini-N measurements in 3 optical bands, however, disfavors a thermal nature of the optical emission. We hence conclude that at least at $t\geq 3$ day after the burst the optical emission of GRB 080503 should be dominated by a non-thermal radiation component. Below let us focus on the spectral properties at ``earlier" times. 

\begin{figure}[!h]
\centering
\includegraphics[width=\columnwidth]{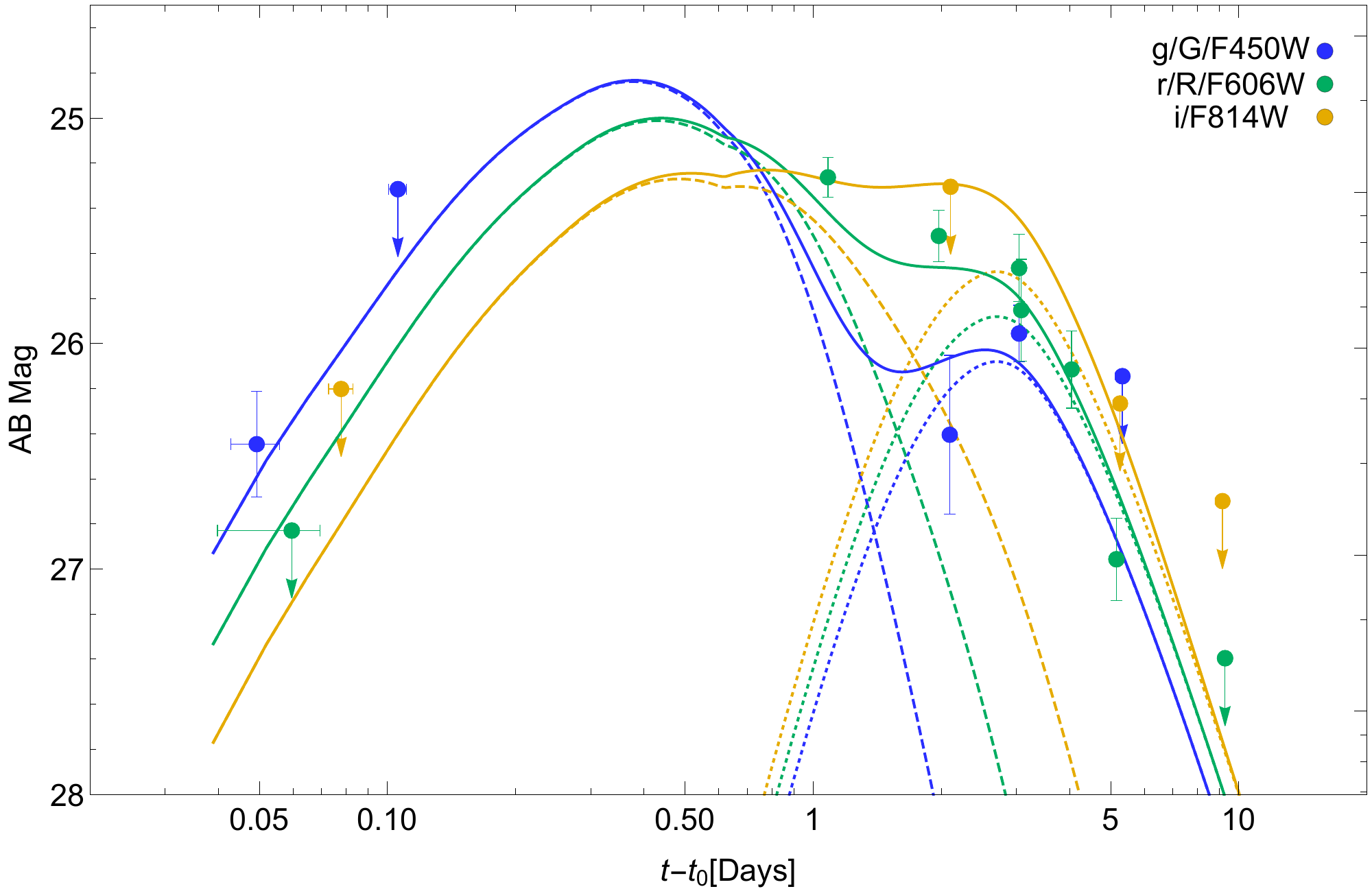}
\caption{
{\bf The multi-band optical emission of GRB 080503 and the interpretation.} For clarity, some observations resulting in loose upper limits are not shown in the plot.
The dashed lines represent the light curves of a blue kilonova with the total mass of the sub-relativistic ejecta of 0.035$M_{\odot}$. The dotted lines represent light curves of forward shock emission of a narrowly jetted outflow with a moderate initial Lorentz factor of $\sim 33$ and a very low number density  of the surrounding medium ($\sim 4.5\times 10^{-7}~{\rm cm^{-3}}$, which is anyhow consistent with the lack of the detection of the host galaxy for this event and hence a possibility of being outside of the galaxy). 
}\label{fig:LC-Model}
\end{figure}

At $t\sim 0.05$ day, there was a detection by Gemini-N in $g$ band. Shortly before and after that time, there were two exposures in $r-$band. No emission was detected and the combined constraint is quite tight, which is actually below the measured $g-$band flux. Such a spectrum is quite unusual and in the synchrotron radiation model a $f_\nu \propto \nu^{1/3}$ is only possible for $\nu_{\rm g}\leq \min\{\nu_{\rm c},~\nu_{\rm m}\}$ unless we assume that the synchrotron-self absorption plays an important role in shaping the optical emission, which is however very unlikely for a reasonable number density of the surrounding medium, where $\nu_{\rm c}$ and $\nu_{\rm m}$ are the cooling frequency and the typical synchrotron radiation frequency \citep{2004RvMP...76.1143P}, respectively.
However, in the forward shock model, $\nu_{\rm m}\approx 5\times 10^{12}~{\rm Hz}~(1+z)^{1/2}E_{\rm k,51}^{1/2}\epsilon_{\rm B,-2}^{1/2}\epsilon_{\rm e,-1}^2[13(p-2)/3(p-1)]^2(t/0.05~{\rm d})^{-3/2}$, where $E_{\rm k}$ is the isotropic equivalent kinetic energy of the outflow, $\epsilon_{\rm e}$ ($\epsilon_{\rm B}$) is the fraction of the forward shock energy converted into the energy of the accelerated electrons (magnetic fields), $z$ is the redshift of the source, $p$ is the power-law index of the electron energy distribution spectral index, and throughout this work we define $Q_{\rm x,n}=Q_{\rm x}/10^{\rm n}$ unless specifically mentioned. One can see that $\nu_{\rm m}(t=0.05~{\rm d})<\nu_{\rm g}\sim 6\times 10^{14}$ Hz unless $\epsilon_{\rm e}\approx 1$.
Even with such an extremely high $\epsilon_{\rm e}$, the problem is not completely solved. This is because to keep the lightcurve rising until $t\sim 1$ day, we actually need $\nu_{\rm m}(t=1~{\rm d})\approx \nu_{\rm g}$, which is impossible unless we artificially assume that almost all the shock energy had been consumed to accelerate just a fraction $\sim 0.1$ of the total electrons (Usually it is assumed that all the electrons have been accelerated by the shock). Even with such an ad hoc assumption, the problem is unresolved since it would predict bright and quickly-decaying X-ray emission and is hence inconsistent with the data shown in \cite{2009ApJ...696.1871P}. We therefore conclude that the synchrotron radiation scenario is disfavored. A more natural interpretation is that the rather hard optical emission is actually the ``intermediate" energy part of a thermal component. As shown in the left panel of Fig.\ref{fig:SED005},  the $g-$band measurement and $r-$band upper limit at $t\sim 0.05$ day can be reproduced with a temperature of $T\sim 2.4\times 10^{4}~(1+z)$ K or even higher. This is certainly possible since AT2017gfo already had a temperature of $\sim 10^{4}$ K at $t\sim 0.5$ day \citep{2017Sci...358.1559K} and a kilonova at early times could be bluer.   
At $t\approx 2$ day after the burst, the source was measured by Gemini-N in $r$ and $g$ bands. Moreover, in this work we also analyzed the simultaneous observation data in $i$ band. Though not detected in such a band, the upper limit still plays an important role in bounding the spectrum. As shown in Fig.\ref{fig:SED005} (the left panel), the spectrum could be fitted with a temperature of $T\sim 5\times 10^{3}~(1+z)$ K, which is higher than that measured in AT2017gfo at the same epoch. Indeed, our $r$-band decline is much shallower than that of AT2017gfo, which naturally suggest a higher temperature unless there is an emerging new radiation component. Anyhow, we would like to comment on the other possibility that the spectrum at $t\sim 2$ day is actually non-thermal. Though the fit of a $\nu^{-0.7}$ is poorer than the thermal spectrum (see the left panel of Fig.\ref{fig:SED005}), it can not be convincingly rejected because of the large uncertainty of the $g$-band measurement. 
Hence, we suggest that the kilonova emission dominates the optical afterglow at most for  $t\leq 2$ d. This is {\it different} from the previous literature which attribute the whole optical emission to the kilonova (or the magnetar boosted macronova).

\section{Discussion}\label{Sec:Discussion}
The current data are insufficient to yield a reliable lightcurve for the kilonova component.With the assumption that the kilonova of GRB 080503 resembles AT2017gfo, we found that if AT2017gfo happened at a redshift of $\sim$ 0.3, its observed $r$-band luminosity is simiar to that of GRB 080503 at $\sim$ 1 day. However, the $r$-band emission of AT2017gfo dropped with time much quicker than the optical emission of GRB 080503 and thus calls for a new radiation component emerging at $t>1$ d.

If the kilonova is dimmer/redder than AT2017gfo, which could be the case because of the large error of the $g$-band measurement at $t\approx 2$ d and hence the sizeable uncertainty about the spectrum. The late time non-thermal radiation would hence be brighter than that in the first scenario and the interpretation is more flexible. As for the physical origin of the non-thermal radiation, the most straightforward speculation is the emerging forward shock emission. Since GRB 080503 has a low gamma-ray luminosity of $L_\gamma \sim 10^{48}$ erg, the empirical $\Gamma_{0}-L_\gamma$ correlation \citep{2012ApJ...751...49L,2012ApJ...755L...6F} would suggest a low bulk Lorentz factor $\Gamma_{0} \sim 25$. This low Lorentz factor is surprisingly well consistent with the value of $\sim 30$  \citep{2009ApJ...696.1871P} needed to account for the peak time $t\sim$ 1 day with a low ambient medium density $n_{0}\sim10^{-6}$ cm$^{-3}$ (constrained with the absence of early afterglow). Though encouraging, the fit to the lightcurve needs some extreme parameters.

We find out that a very narrow jet opening angle is required. This is because the optical radiation component should rise rapidly (quicker than $t$) and then decline quickly (rapider than $t^{-1.6}$), together with a sharp transition. In the standard fireball external shock model, such a feature can only be achieved with a narrow jet with a half-opening angle of $\theta_{\rm j}\leq 1/\Gamma_{0}$, for which the quick rise is due to the rapidly increasing number of electrons involved in the radiation and the rapid decline is due to the jet effect. The late time optical emission can be roughly reproduced by the following parameters, including $\Gamma_0=33$, $\epsilon_{\rm e}=0.45$, $\epsilon_{\rm B}=0.1$, $n_0=5\times 10^{-7}~{\rm cm^{-3}}$, $E_{\rm k}=7\times 10^{51} {\rm erg}$, $\theta_j=0.021$, $\theta_v=0$ and $p=2.4$ (see Fig.\ref{fig:LC-Model} for the numerical result calculated with \cite{2006MNRAS.369..197F}). The extremely low medium density $n_0$ is consistent with the constraint given by \cite{2009ApJ...696.1871P} and the very low equivalent Hydrogen column density of the possible host galaxy of GRB 080503 derived by X-ray data.

The early time kilonova emission was calculated with a spherical symmetry two-component model \citep{2020ApJ...891..152H}, including a high $\rm Y_{e}$ part (i.e., almost lanthanide-free) and a low $\rm Y_{e}$ part. The total mass of the sub-relativistic ejecta is 0.035$M_{\odot}$, and the density profile is set to be $\rho(t,v)\propto (v_{\rm ej}/0.1c)^{-4.5}$ for $0.1c\leq v_{\rm ej} \leq 0.4c$, where $c$ is the speed of light in the vacuum. Such a narrow $\theta_{\rm j}$, though unusual, is still possible, as found in  GRB 090510 and GRB 061201, and possibly also GRB 160821B \citep{2019ApJ...883...48L}. Here we would like to remind that the request of a very narrow jet can be released if just the optical emission at $t\sim 0.05$ day was dominated by the kilonova, for which the forward shock emission could peak at $t<1$ day and the transition to the quick decline phase would be more smooth and natural. 

\begin{figure}[h]
    \centering
    \includegraphics{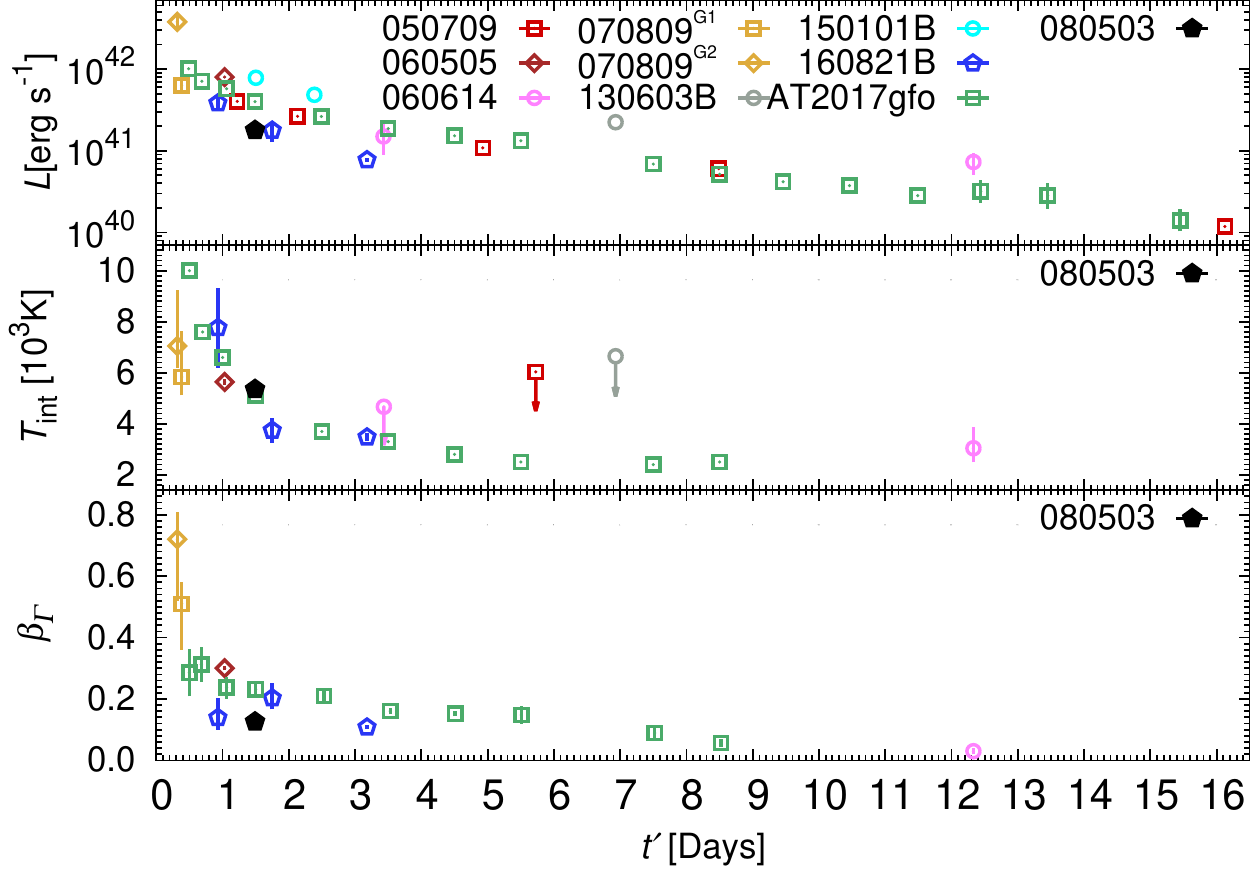}
    \caption{{\bf Comparison of properties of kilonova.} Assuming the redshift of GRB 080503 is $\sim$ 0.3, physical properties of possible kilonova of GRB 080503 is consistent with that of 160821B. Data points of GRB 060505 are taken from \cite{2021arXiv210907694J}, while all other data points are taken from \cite{2020NatAs...4...77J}. The G1 and G2 mark for GRB 070809 represent different redshifts of 0.22 and 0.47, respectively.}
    \label{fig:KN_compare}
\end{figure}

Assuming that the redshift of GRB 080503 is 0.3, the luminosity $L$ and the intrinsic temperature $T_{\rm int}$ of kilonova model are $\sim 1.79\times10^{41}$ erg/s and $4115\times(1+0.3)=5350$ K at $t'=1.965/(1+0.3)=1.49$ day in GRB rest frame, respectively. Note that at such epoch, afterglow became bright enough to be comparable with kilonova, which means that the SED at 1.49 day is a mixture of power-law and black-body components, hence the temperature of kilonova derived in model calculation is lower than that estimated by a single black-body SED (i.e., $\sim 5000(1+z)$ show in left panel of Figure \ref{fig:SED005}). The velocity of outflow in unit of light speed $\beta_{\Gamma}$ can be derived with $(1+\beta_{\Gamma})\beta_{\Gamma}\Gamma\approx0.4L^{1/2}_{42}(T_{\rm int}/6000{\rm K})^{-2}(t'/1{\rm day})^{-1}$\citep{2020NatAs...4...77J}, where $\Gamma=1/\sqrt{1-\beta_{\Gamma}^2}$ and $L_{42}$ is luminosity in unit of $10^{42}$ erg/s. Figure \ref{fig:KN_compare} shows properties of kilonova component of GRB 080503 and other kilonova or candidates. The intrinsic temperature of kilonova of GRB 080503 is similar to others, but the luminosity and outflow velocity is somehow lower than others except the kilonova of GRB 160821B. Anyway, such a case is reasonable since the initial bulk Lorentz factor of afterglow is also very low $\sim33$. All signals of kilonova, afterglow and initial X-ray and $\gamma$-ray show that GRB 080503 could be an extremely weak GRB, which happened outside of its host galaxy.

Though the current optical data can not pin down when the thermal-like emission  became subdominant in the time range of $0.05-3$ day, it is very likely that the first epoch data with an unusual hard spectrum was dominated by a blue kilonova at an observed temperature of $\sim 2.4\times 10^{4}$ K. If correct, this would be the earliest detection of the kilonova up to date. Previously this record was held by AT2017gfo \citep{2017Natur.551...64A} and then updated, though slightly, by GRB 070809 \citep{2020NatAs...4...77J}. The very early emission phase of the kilonova is expected to shed valuable light on the composition of the sub-relativistic outflow launched by the neutron star merger. For both Type I (expected to be popular in the next decade) and Type IV lightcurves of optical transients powered by the neutron star mergers, thanks to the increasing accuracy of the localization of the gravitational wave events by the LIGO/Virgo/KAGRA network \citep{2020LRR....23....3A}, kilonova events are expected to be detected in its very early phase. Hence, early temporal and spectral evolution of kilonova will be well measured, with which the r-process nuclear synthesis in different mergers/remnant scenarios will be further revealed \citep{2019LRR....23....1M}.

\section{Acknowledgement}
This work was supported in part by NSFC under grants of No. 12225305, No. 11921003, No. 12233011, No. 11933010 and No. 12073080, the China Manned Space Project (NO.CMS-CSST-2021-A13), Major Science and Technology Project of Qinghai Province  (2019-ZJ-A10), Key Research Program of Frontier Sciences (No. QYZDJ-SSW-SYS024). SC has been supported by ASI grant I/004/11/0.

This research made use of Photutils, an Astropy package for detection and photometry of astronomical sources \citep{larry_bradley_2022_6825092}.

\newpage
\begin{table}[h]
\centering
\begin{tabular}{ccccccc}
\hline
\hline
Time & Exposure & Instrument & Filter & $\lambda^a$ & Magnitude$^b$ & Flux \\
(Days) & (s) & & & (\r{A}) & (AB) & ($\mu$Jy) \\
\hline
\hline
0.049 & 900 & GMOS/Gemini-N & $g$ & 4843.7 & 26.45$\pm$0.23 & 0.096$\pm$0.021 \\
0.051 & 90 & LRIS/Keck-I & $B$ & 4369.9 & $>$25.00 & $<$0.362 \\
0.059 & 600 & LRIS/Keck-I & $R$ & 6417.0 & $>$25.79 & $<$0.176 \\
0.060 & 1080$^c$ & GMOS/Gemini-N & $r$ & 6314.3 & $>$26.83 & $<$0.067 \\
0.078 & 720 & GMOS/Gemini-N & $i$ & 7716.9 & $>$26.20 & $<$0.120 \\
0.093 & 900 & GMOS/Gemini-N & $z$ & 9033.3 & $>$25.36 & $<$0.261 \\
0.106 & 720 & GMOS/Gemini-N & $g$ & 4843.7 & $>$25.32 & $<$0.271 \\
1.084 & 1800 & GMOS/Gemini-N & $r$ & 6314.3 & 25.26$\pm$0.09 & 0.285$\pm$0.023 \\
1.974 & 1620 & GMOS/Gemini-N & $r$ & 6314.3 & 25.52$\pm$0.11 & 0.224$\pm$0.023 \\
2.095 & 720 & GMOS/Gemini-N & $g$ & 4843.7 & 26.41$\pm$0.35 & 0.099$\pm$0.032 \\
2.105$^d$ & 720 & GMOS/Gemini-N & $i$ & 7716.9 & $>$25.31 & $<$0.274 \\
3.048$^d$ & 6000 & LRIS/Keck-I & $G$ & 4730.5 & 25.96$\pm$0.13 & 0.150$\pm$0.017 \\
3.048$^d$ & 6000 & LRIS/Keck-I & $R$ & 6417.0 & 25.67$\pm$0.15 & 0.197$\pm$0.027 \\
3.084 & 2160 & GMOS/Gemini-N & $r$ & 6314.3 & 25.85$\pm$0.23 & 0.166$\pm$0.034 \\
4.050 & 2880 & GMOS/Gemini-N & $r$ & 6314.3 & 26.12$\pm$0.17 & 0.130$\pm$0.020 \\
5.162 & 4600 & WFPC2/HST & F606W & 5996.8 & 26.96$\pm$0.18 & 0.060$\pm$0.010 \\
5.262 & 2100 & WFPC2/HST & F814W & 8012.2 & $>$26.27 & $<$0.113 \\
5.328 & 2100 & WFPC2/HST & F450W & 4555.4 & $>$26.15 & $<$0.126 \\
9.157 & 4000 & WFPC2/HST & F814W & 8012.2 & $>$26.70 & $<$0.076 \\
9.288 & 4000 & WFPC2/HST & F606W & 5996.8 & $>$27.40 & $<$0.040 \\
\hline
\end{tabular} \\
$a$. Effective wavelengths of Gemini's filters are derived by the correlation curve of filter transmission with CCD quantum efficiency, assuming a flat spectrum. While effective wavelength of Keck's and HST's filters are taken from Keck LRIS webpage and WFPC2 Instrument Handbook. \\
$b$. For LRIS/Keck-I, 5-$\sigma$ upper-limits are reported due to its poor image quality compared with other instruments, while 3-$\sigma$ upper-limits are reported for GMOS/Gemini-N and WFPC2/HST. \\
$c$. Here we combine the 180s exposure centered at $0.0408$ day and the 900s exposure centered at $0.0625$ day after the trigger of GRB 080503. \\
$d$. These data points are reported for the first time.
\caption{{\bf Photometric observations of GRB080503.} Milky Way extinction correction has been applied with $E(B-V)=0.0525$ \cite{2011ApJ...737..103S}, $A_{\rm g}=0.1896$, $A_{\rm r}=0.1400$, $A_{\rm i}=0.1043$, $A_{\rm z}=0.0749$, $A_{\rm B}=0.2174$, $A_{\rm G}=0.1964$, $A_{\rm R}=0.1368$, $A_{\rm F450W}=0.2063$, $A_{\rm F606W}=0.1475$, $A_{\rm F814W}=0.0973$. All are in units of magnitude.}
\label{tab:080503Phot}
\end{table}

\newpage
\clearpage
\begin{table}[h]
    \centering
    \begin{tabular}{cc}
    \hline
    \hline
    Energy & Flux \\
    (keV) & (nJy) \\
    \hline
    \hline
    0.67$\pm$0.22 & 1.51$\pm$0.49 \\
    1.12$\pm$0.23 & 1.02$\pm$0.32 \\
    1.84$\pm$0.50 & 0.57$\pm$0.20 \\
    2.88$\pm$0.54 & 1.21$\pm$0.43 \\
    5.65$\pm$2.23 & 0.36$\pm$0.18 \\
    \hline
    \end{tabular}
    \caption{{\bf Intrinsic(Un-absorbed) X-ray Spectrum of GRB 080503 at $\sim$4.5 day.} The absorption is dominated by Milky Way with equivalent Hydrogen column density $N_{\rm H}=6.99\times10^{20}{\rm cm}^{-2}$, see Table \ref{tab:X-ray_Fitting}. All these data are plotted in right panel of Figure \ref{fig:SED005}.}
\label{tab:X-ray_Spec}
\end{table}

\newpage
\clearpage
\begin{table}[h]
\centering 
\begin{tabular}{ccc} 
\hline
\hline
parameters & value$^a$ & unit \\
\hline
\hline
${N_{\rm H}}^b$ &  $6.99\times10^{-2}$ & $10^{22}$ cm$^{-2}$ \\
\textit{flux} (0.3 - 10 keV) & $(1.22_{-0.208}^{+0.355})\times10^{-14}$ & $\rm ergs/cm^2/s$ \\
\textit{$\beta$} & $0.73\pm0.27$ & / \\
\hline
\end{tabular} \\
$a$. Uncertainties are reported as 68.3 $\%$ confidence interval (i.e., 1 $\sigma$ uncertainty). The asymmetric uncertainty is reported as the median with a lower bound reaches a quantile of $15.87\%$ and an upper bound reaches $87.13\%$. \\
$b$. The equivalent Hydrogen column density of Milky Way, $N_{\rm H}$, is fixed as same as the value announced before (see \url{https://www.swift.ac.uk/xrt\_spectra/00310785/}). The intrinsic absorption is so small $\sim 5.82\times10^{-9}$ that be ignored by comparison.
\caption{{\bf The best-fit parameters of Chandra X-ray spectrum.} The fitted spectral model is \textit{phabs*zphabs*cflux*powerlaw} , where \textit{phabs} describes the absorption of Milky Way and \textit{zphabs} fits the equivalent Hydrogen column density that caused spectral absorption by both of intrinsic(i.e., the host galaxy) and intergalactic matter with consideration of cosmological redshift. \textit{cflux} is a mark to tell \textit{X-Spec} to calculate the flux of the power-law spectrum. The redshift is $z=0.3$ in the model. $Cash$ statistics method was applied to fit the Chandra X-ray spectrum data of GRB 080503.}
\label{tab:X-ray_Fitting}
\end{table}

\newpage
\clearpage
\begin{table}[h]
\centering
\begin{tabular}{cccccc}
\hline
\hline
RA & DEC & $g$ & $r$ & $B^a$ & $R^a$ \\
(J2000) & (J2000) & (AB) & (AB) & (Vega) & (Vega) \\
\hline
\hline
19:05:54.243 & +68:47:14.40 & 20.46(0.051) & 19.79(0.037) &      /       & 19.56(0.088) \\
19:05:58.596 & +68:47:47.03 & 21.83(0.054) & 20.53(0.037) & 22.61(0.062) & 20.21(0.092) \\
19:05:58.755 & +68:47:43.14 & 20.98(0.051) & 19.67(0.037) & 21.77(0.059) & 19.34(0.088) \\
19:06:00.765 & +68:48:20.36 & 19.95(0.051) & 19.65(0.037) & 20.27(0.058) & 19.46(0.087) \\
19:06:02.468 & +68:48:07.02 & 22.78(0.067) & 21.34(0.040) &      /       & 21.00(0.111) \\
19:06:05.033 & +68:48:15.81 & 21.53(0.052) & 20.21(0.037) & 22.32(0.060) & 19.89(0.090) \\
19:06:06.768 & +68:47:06.96 & 22.71(0.065) & 21.27(0.039) &      /       & 20.92(0.108) \\
19:06:07.447 & +68:47:20.58 & 21.21(0.051) & 20.89(0.038) & 21.54(0.060) & 20.70(0.089) \\
19:06:08.946 & +68:49:52.82 & 18.77(0.050) & 18.02(0.037) & 19.30(0.058) &      /       \\
19:06:10.223 & +68:48:26.38 & 20.07(0.051) & 18.98(0.037) & 20.76(0.058) & 18.69(0.087) \\
19:06:16.303 & +68:46:41.09 & 19.81(0.050) & 18.62(0.037) & 20.55(0.058) & 18.31(0.087) \\
19:06:20.587 & +68:44:58.47 & 21.52(0.052) & 20.35(0.037) & 22.23(0.060) &      /       \\
19:06:21.636 & +68:49:33.24 & 20.81(0.051) & 19.60(0.037) & 21.55(0.059) & 19.29(0.088) \\
19:06:22.926 & +68:45:45.37 & 21.79(0.053) & 20.39(0.037) & 22.62(0.062) & 20.05(0.092) \\
19:06:24.810 & +68:48:47.65 & 18.50(0.050) & 17.85(0.037) & 18.98(0.058) &      /       \\
19:06:25.325 & +68:47:17.82 & 21.47(0.052) & 20.35(0.037) & 22.17(0.060) & 20.05(0.090) \\
19:06:25.858 & +68:46:57.49 & 23.08(0.077) & 21.78(0.043) &      /       & 21.46(0.126) \\
19:06:27.123 & +68:45:04.09 & 21.52(0.052) & 20.37(0.037) & 22.24(0.060) &      /       \\
19:06:27.450 & +68:46:55.32 & 21.22(0.051) & 20.69(0.038) & 21.63(0.060) & 20.48(0.089) \\
19:06:28.758 & +68:46:09.71 & 21.97(0.055) & 20.65(0.038) &      /       & 20.32(0.093) \\
19:06:28.816 & +68:49:02.35 & 21.20(0.051) & 20.01(0.037) & 21.93(0.060) & 19.70(0.089) \\
19:06:31.165 & +68:45:59.64 & 20.90(0.051) & 20.40(0.037) & 21.31(0.059) &      /       \\
19:06:31.174 & +68:48:31.88 & 19.09(0.050) & 17.80(0.037) & 19.87(0.058) &      /       \\
19:06:31.794 & +68:49:36.52 & 20.25(0.051) & 19.03(0.037) & 20.99(0.058) & 18.72(0.087) \\
19:06:32.698 & +68:46:33.02 & 20.76(0.051) & 20.25(0.037) & 21.17(0.059) & 20.04(0.088) \\
19:06:32.819 & +68:48:01.52 & 22.05(0.055) & 20.66(0.038) &      /       & 20.32(0.094) \\
19:06:33.717 & +68:46:35.32 & 19.71(0.050) & 19.12(0.037) & 20.16(0.058) & 18.89(0.087) \\
19:06:37.493 & +68:49:01.79 & 20.94(0.051) & 19.65(0.037) & 21.71(0.059) & 19.33(0.088) \\
19:06:38.455 & +68:47:44.42 & 19.76(0.050) & 19.34(0.037) & 20.12(0.058) & 19.14(0.087) \\
19:06:38.785 & +68:47:47.76 & 19.79(0.050) & 19.01(0.037) & 20.33(0.058) & 18.76(0.087) \\
19:06:41.849 & +68:48:04.92 & 18.79(0.050) & 18.39(0.037) & 19.16(0.058) & 18.19(0.087) \\
19:06:42.416 & +68:48:08.44 & 21.59(0.053) & 21.03(0.038) & 22.02(0.061) & 20.81(0.091) \\
\hline
\end{tabular} \\
$a$. Results derived from formula in \cite{2002AJ....123.2121S}. To get AB magnitudes, -0.128 and 0.178 should be added for $B$ and $R$ bands, respectively.
\caption{{\bf Secondary standard stars in Gemini and Keck field on 3 May 2008.} Values in parentheses are 1$\sigma$ error of measurements and elements marked with '/' represent these stars are not used to calibrate corresponding zeropoints, due to that they are saturated in Keck's images, or just are outside of images either of Keck or Gemini.}
\label{tab:GandKMay3}
\end{table}

\newpage
\clearpage
\begin{table}[h]
\centering
\begin{tabular}{cccccc}
\hline
\hline
RA & DEC & $g$ & $r$ & $G^a$ & $R$ \\
(J2000) & (J2000) & (AB) & (AB) & (AB) & (Vega) \\
\hline
\hline
19:06:00.765 & +68:48:20.36 & 19.95(0.051) & 19.65(0.037) & 19.95(0.051) & 19.46(0.087) \\
19:06:02.468 & +68:48:07.02 & 22.78(0.067) & 21.34(0.040) &      /       & 21.00(0.111) \\
19:06:05.033 & +68:48:15.81 & 21.53(0.052) & 20.21(0.037) & 21.53(0.052) & 19.89(0.090) \\
19:06:07.447 & +68:47:20.58 & 21.21(0.051) &      /       & 21.21(0.051) &      /       \\
19:06:10.223 & +68:48:26.38 & 20.07(0.051) &      /       & 20.07(0.051) &      /       \\
19:06:21.636 & +68:49:33.24 & 20.81(0.051) &      /       & 20.81(0.051) &      /       \\
19:06:22.926 & +68:45:45.37 & 21.79(0.053) & 20.39(0.037) & 21.79(0.053) & 20.05(0.092) \\
19:06:25.325 & +68:47:17.82 & 21.47(0.052) & 20.35(0.037) & 21.47(0.052) & 20.05(0.090) \\
19:06:25.858 & +68:46:57.49 & 23.08(0.077) & 21.78(0.043) &      /       & 21.46(0.126) \\
19:06:27.123 & +68:45:04.09 & 21.52(0.052) & 20.37(0.037) &      /       & 20.07(0.090) \\
19:06:27.450 & +68:46:55.32 & 21.22(0.051) & 20.69(0.038) &      /       & 20.48(0.089) \\
19:06:27.545 & +68:49:18.60 & 22.11(0.056) & 21.02(0.038) &      /       & 20.72(0.095) \\
19:06:28.758 & +68:46:09.71 & 21.97(0.055) & 20.65(0.038) & 21.97(0.055) & 20.32(0.093) \\
19:06:28.816 & +68:49:02.35 & 21.20(0.051) & 20.01(0.037) & 21.20(0.051) & 19.70(0.089) \\
19:06:29.024 & +68:47:49.73 & 23.13(0.079) & 21.78(0.043) &      /       & 21.45(0.130) \\
19:06:29.083 & +68:46:12.50 & 23.01(0.074) & 21.86(0.044) &      /       & 21.56(0.123) \\
19:06:30.354 & +68:48:06.11 & 24.15(0.163) & 22.57(0.059) &      /       & 22.21(0.255) \\
19:06:31.165 & +68:45:59.64 & 20.90(0.051) & 20.40(0.037) & 20.90(0.051) & 20.19(0.088) \\
19:06:31.174 & +68:48:31.88 & 19.09(0.050) &      /       & 19.09(0.050) &      /       \\
19:06:31.678 & +68:49:44.83 & 22.32(0.058) & 21.46(0.040) &      /       & 21.20(0.100) \\
19:06:31.794 & +68:49:36.52 & 20.25(0.051) &      /       & 20.25(0.051) &      /       \\
19:06:32.506 & +68:50:00.68 & 23.54(0.101) & 22.13(0.048) &      /       & 21.80(0.163) \\
19:06:32.698 & +68:46:33.02 & 20.76(0.051) &      /       & 20.76(0.051) &      /       \\
19:06:32.808 & +68:45:56.80 & 20.13(0.051) &      /       & 20.13(0.051) &      /       \\
19:06:32.819 & +68:48:01.52 & 22.05(0.055) & 20.66(0.038) & 22.05(0.055) & 20.32(0.094) \\
19:06:33.717 & +68:46:35.32 & 19.71(0.050) & 19.12(0.037) & 19.71(0.050) & 18.89(0.087) \\
19:06:34.361 & +68:46:22.52 & 22.24(0.057) & 21.04(0.038) &      /       & 20.73(0.097) \\
19:06:35.994 & +68:46:11.50 & 20.85(0.051) & 20.31(0.037) & 20.85(0.051) & 20.09(0.088) \\
19:06:37.493 & +68:49:01.79 & 20.94(0.051) & 19.65(0.037) & 20.94(0.051) & 19.33(0.088) \\
19:06:38.455 & +68:47:44.42 & 19.76(0.050) &      /       & 19.76(0.050) &      /       \\
19:06:38.785 & +68:47:47.76 & 19.79(0.050) &      /       & 19.79(0.050) &      /       \\
19:06:40.315 & +68:47:13.94 & 20.80(0.051) &      /       & 20.80(0.051) &      /       \\
19:06:41.704 & +68:46:18.30 & 20.95(0.051) & 19.78(0.037) & 20.95(0.051) & 19.48(0.088) \\
19:06:41.849 & +68:48:04.92 & 18.79(0.050) &      /       & 18.79(0.050) &      /       \\
19:06:42.416 & +68:48:08.44 & 21.59(0.053) & 21.03(0.038) & 21.59(0.053) & 20.81(0.091) \\
19:06:46.069 & +68:46:07.10 & 20.12(0.051) &      /       & 20.12(0.051) &      /       \\
19:06:46.621 & +68:47:43.83 & 20.30(0.051) &      /       & 20.30(0.051) &      /       \\
19:06:47.632 & +68:47:16.52 & 19.93(0.051) &      /       & 19.93(0.051) &      /       \\
19:06:54.422 & +68:46:56.84 & 19.88(0.050) & 19.53(0.037) &      /       & 19.34(0.087) \\
\hline
\end{tabular} \\
$a$. Effective wavelength of g and G are similar, hence we use g-band AB magnitudes as inferred G-band AB magnitudes.
\caption{{\bf Secondary standard stars in Gemini and Keck field on May 6.} Same as Table \ref{tab:GandKMay3}.}
\label{tab:GandKMay6}
\end{table}

\bibliography{sample631}{}
\bibliographystyle{aasjournal}

\end{document}